\documentclass{svmult}
\usepackage{epsfig,graphicx} 
\usepackage[bottom]{footmisc}
\usepackage{natbib,url}      
\bibpunct{(}{)}{;}{a}{}{,}   
\def\cite#1{\citealp{#1}}    
\def\authorindex#1{}  

\begin{document}\newcount\preprintheader\preprintheader=1


\def\thisvolume{these proceedings}

\def\aj{{AJ}}			
\def\araa{{ARA\&A}}		
\def\apj{{ApJ}}			
\def\apjl{{ApJ}}		
\def\apjs{{ApJS}}		
\def\ao{{Appl.\ Optics}} 
\def\apss{{Ap\&SS}}		
\def\aap{{A\&A}}		
\def\aapr{{A\&A~Rev.}}		
\def\aaps{{A\&AS}}		
\def\an{{Astron.\ Nachrichten}}
\def\aspcs{{ASP Conf.\ Ser.}}
\def\assp{{Astrophys.\ \& Space Sci.\ Procs., Springer, Heidelberg}}
\def\azh{{AZh}}			
\def\baas{{BAAS}}		
\def\jrasc{{JRASC}}	
\def\memras{{MmRAS}}		
\def\mnras{{MNRAS}}
\def\nat{{Nat}}		
\def\pra{{Phys.\ Rev.\ A}} 
\def\prb{{Phys.\ Rev.\ B}}		
\def\prc{{Phys.\ Rev.\ C}}		
\def\prd{{Phys.\ Rev.\ D}}		
\def\prl{{Phys.\ Rev.\ Lett.}} 
\def\pasp{{PASP}}
\def\pasj{{PASJ}}		
\def\qjras{{QJRAS}}
\def\science{{Sci}}		
\def\skytel{{S\&T}}		
\def\solphys{{Solar\ Phys.}} 
\def\sovast{{Soviet\ Ast.}}  
\def\ssr{{Space\ Sci.\ Rev.}}
\def\svassp{{Astrophys.\ Space Sci.\ Procs., Springer, Heidelberg}}
\def\zap{{ZAp}}			
\let\astap=\aap
\let\apjlett=\apjl
\let\apjsupp=\apjs
\def\grl{{Geophys.\ Res.\ Lett.}}  
\def\jgr{{J. Geophys.\ Res.}} 

\def\ion#1#2{{\rm #1}\,{\uppercase{#2}}}  
\def\deg{\hbox{$^\circ$}}
\def\sun{\hbox{$\odot$}}
\def\earth{\hbox{$\oplus$}}
\def\la{\mathrel{\hbox{\rlap{\hbox{\lower4pt\hbox{$\sim$}}}\hbox{$<$}}}}
\def\ga{\mathrel{\hbox{\rlap{\hbox{\lower4pt\hbox{$\sim$}}}\hbox{$>$}}}}
\def\sq{\hbox{\rlap{$\sqcap$}$\sqcup$}}
\def\arcmin{\hbox{$^\prime$}}
\def\arcsec{\hbox{$^{\prime\prime}$}}
\def\fd{\hbox{$.\!\!^{\rm d}$}}
\def\fh{\hbox{$.\!\!^{\rm h}$}}
\def\fm{\hbox{$.\!\!^{\rm m}$}}
\def\fs{\hbox{$.\!\!^{\rm s}$}}
\def\fdg{\hbox{$.\!\!^\circ$}}
\def\farcm{\hbox{$.\mkern-4mu^\prime$}}
\def\farcs{\hbox{$.\!\!^{\prime\prime}$}}
\def\fp{\hbox{$.\!\!^{\scriptscriptstyle\rm p}$}}
\def\micron{\hbox{$\mu$m}}
\def\onehalf{\hbox{$\,^1\!/_2$}}	
\def\onethird{\hbox{$\,^1\!/_3$}}
\def\twothirds{\hbox{$\,^2\!/_3$}}
\def\onequarter{\hbox{$\,^1\!/_4$}}
\def\threequarters{\hbox{$\,^3\!/_4$}}
\def\ubv{\hbox{$U\!BV$}}		
\def\ubvr{\hbox{$U\!BV\!R$}}		
\def\ubvri{\hbox{$U\!BV\!RI$}}		
\def\ubvrij{\hbox{$U\!BV\!RI\!J$}}		
\def\ubvrijh{\hbox{$U\!BV\!RI\!J\!H$}}		
\def\ubvrijhk{\hbox{$U\!BV\!RI\!J\!H\!K$}}		
\def\ub{\hbox{$U\!-\!B$}}		
\def\bv{\hbox{$B\!-\!V$}}		
\def\vr{\hbox{$V\!-\!R$}}		
\def\ur{\hbox{$U\!-\!R$}}


\def\labelitemi{{\bf --}}  

\def\rmit#1{{\it #1}}              
\def\rmit#1{{\rm #1}}              
\def\etal{\rmit{et al.}}           
\def\etc{\rmit{etc.}}           
\def\ie{\rmit{i.e.,}}              
\def\eg{\rmit{e.g.,}}              
\def\cf{cf.}                       
\def\viz{\rmit{viz.}}
\def\vs{\rmit{vs.}}

\def\rot{\hbox{\rm rot}}
\def\div{\hbox{\rm div}}
\def\lesssim{\mathrel{\hbox{\rlap{\hbox{\lower4pt\hbox{$\sim$}}}\hbox{$<$}}}}
\def\gtrsim{\mathrel{\hbox{\rlap{\hbox{\lower4pt\hbox{$\sim$}}}\hbox{$>$}}}}
\def\mathstacksym#1#2#3#4#5{\def#1{\mathrel{\hbox to 0pt{\lower 
    #5\hbox{#3}\hss} \raise #4\hbox{#2}}}}
\mathstacksym\lesssim{$<$}{$\sim$}{1.5pt}{3.5pt} 
\mathstacksym\gtrsim{$>$}{$\sim$}{1.5pt}{3.5pt} 
\mathstacksym\lrarrow{$\leftarrow$}{$\rightarrow$}{2pt}{1pt} 
\mathstacksym\lessgreat{$>$}{$<$}{3pt}{3pt} 

\def\dif{\: {\rm d}}                       
\def\ep{\:{\rm e}^}                        
\def\dash{\hbox{$\,-\,$}}                  
\def\is{\!=\!}                             

\def\starname#1#2{${#1}$\,{\rm {#2}}}  
\def\Teff{\hbox{$T_{\rm eff}$}}   

\def\kms{\hbox{km$\;$s$^{-1}$}}
\def\ms{\hbox{m$\;$s$^{-1}$}}
\def\Mxcm{\hbox{Mx\,cm$^{-2}$}}    

\def\Bapp{\hbox{$B_{\rm app}$}}    

\def\komega{($k, \omega$)}                 
\def\kf{($k_h,f$)}                         
\def\VminI{\hbox{$V\!\!-\!\!I$}}           
\def\IminI{\hbox{$I\!\!-\!\!I$}}           
\def\VminV{\hbox{$V\!\!-\!\!V$}}           
\def\Xt{\hbox{$X\!\!-\!t$}}                

\def\level #1 #2#3#4{$#1 \: ^{#2} \mbox{#3} ^{#4}$}   

\def\specchar#1{\uppercase{#1}}    
\def\AlI{\mbox{Al\,\specchar{i}}}  
\def\BI{\mbox{B\,\specchar{i}}} 
\def\BII{\mbox{B\,\specchar{ii}}}  
\def\BaI{\mbox{Ba\,\specchar{i}}}  
\def\BaII{\mbox{Ba\,\specchar{ii}}} 
\def\CI{\mbox{C\,\specchar{i}}} 
\def\CII{\mbox{C\,\specchar{ii}}} 
\def\CIII{\mbox{C\,\specchar{iii}}} 
\def\CIV{\mbox{C\,\specchar{iv}}} 
\def\CaI{\mbox{Ca\,\specchar{i}}} 
\def\CaII{\mbox{Ca\,\specchar{ii}}} 
\def\CaIII{\mbox{Ca\,\specchar{iii}}} 
\def\CoI{\mbox{Co\,\specchar{i}}} 
\def\CrI{\mbox{Cr\,\specchar{i}}} 
\def\CriI{\mbox{Cr\,\specchar{ii}}} 
\def\CsI{\mbox{Cs\,\specchar{i}}} 
\def\CsII{\mbox{Cs\,\specchar{ii}}} 
\def\CuI{\mbox{Cu\,\specchar{i}}} 
\def\FeI{\mbox{Fe\,\specchar{i}}} 
\def\FeII{\mbox{Fe\,\specchar{ii}}} 
\def\FeIX{\mbox{Fe\,\specchar{ix}}}
\def\FeX{\mbox{Fe\,\specchar{x}}}
\def\FeXVI{\mbox{Fe\,\specchar{xvi}}}
\def\FrI{\mbox{Fr\,\specchar{i}}}
\def\HI{\mbox{H\,\specchar{i}}} 
\def\HII{\mbox{H\,\specchar{ii}}} 
\def\Hmin{\hbox{\rmH$^{^{_{\scriptstyle -}}}$}}      
\def\Hemin{\hbox{{\rm He}$^{^{_{\scriptstyle -}}}$}} 
\def\HeI{\mbox{He\,\specchar{i}}} 
\def\HeII{\mbox{He\,\specchar{ii}}} 
\def\HeIII{\mbox{He\,\specchar{iii}}} 
\def\KI{\mbox{K\,\specchar{i}}} 
\def\KII{\mbox{K\,\specchar{ii}}} 
\def\KIII{\mbox{K\,\specchar{iii}}} 
\def\LiI{\mbox{Li\,\specchar{i}}} 
\def\LiII{\mbox{Li\,\specchar{ii}}} 
\def\LiIII{\mbox{Li\,\specchar{iii}}} 
\def\MgI{\mbox{Mg\,\specchar{i}}} 
\def\MgII{\mbox{Mg\,\specchar{ii}}} 
\def\MgIII{\mbox{Mg\,\specchar{iii}}} 
\def\MnI{\mbox{Mn\,\specchar{i}}} 
\def\NI{\mbox{N\,\specchar{i}}}
\def\NIV{\mbox{N\,\specchar{iv}}}
\def\NaI{\mbox{Na\,\specchar{i}}}
\def\NaII{\mbox{Na\,\specchar{ii}}}
\def\NaIII{\mbox{Na\,\specchar{iii}}}
\def\NeVIII{\mbox{Ne\,\specchar{viii}}} 
\def\NiI{\mbox{Ni\,\specchar{i}}} 
\def\NiII{\mbox{Ni\,\specchar{ii}}}
\def\NiIII{\mbox{Ni\,\specchar{iii}}} 
\def\OI{\mbox{O\,\specchar{i}}} 
\def\OVI{\mbox{O\,\specchar{vi}}}
\def\RbI{\mbox{Rb\,\specchar{i}}} 
\def\SII{\mbox{S\,\specchar{ii}}} 
\def\SiI{\mbox{Si\,\specchar{i}}} 
\def\SiII{\mbox{Si\,\specchar{ii}}} 
\def\SrI{\mbox{Sr\,\specchar{i}}}
\def\SrII{\mbox{Sr\,\specchar{ii}}}
\def\TiI{\mbox{Ti\,\specchar{i}}} 
\def\TiII{\mbox{Ti\,\specchar{ii}}} 
\def\TiIII{\mbox{Ti\,\specchar{iii}}} 
\def\TiIV{\mbox{Ti\,\specchar{iv}}} 
\def\VI{\mbox{V\,\specchar{i}}} 
\def\HtwoO{\mbox{H$_2$O}}        
\def\Otwo{\mbox{O$_2$}}          

\def\Halpha{\mbox{H\hspace{0.1ex}$\alpha$}} 
\def\Ha{\mbox{H\hspace{0.2ex}$\alpha$}}
\def\Hbeta{\mbox{H\hspace{0.2ex}$\beta$}}
\def\Hgamma{\mbox{H\hspace{0.2ex}$\gamma$}}
\def\Hdelta{\mbox{H\hspace{0.2ex}$\delta$}}
\def\Hepsilon{\mbox{H\hspace{0.2ex}$\epsilon$}}
\def\Hzeta{\mbox{H\hspace{0.2ex}$\zeta$}}
\def\Lyalpha{\mbox{Ly$\hspace{0.2ex}\alpha$}}
\def\Lybeta{\mbox{Ly$\hspace{0.2ex}\beta$}}
\def\Lygamma{\mbox{Ly$\hspace{0.2ex}\gamma$}}
\def\Lycont{\mbox{Ly\hspace{0.2ex}{\small cont}}}
\def\Baalpha{\mbox{Ba$\hspace{0.2ex}\alpha$}}
\def\Babeta{\mbox{Ba$\hspace{0.2ex}\beta$}}
\def\Bacont{\mbox{Ba\hspace{0.2ex}{\small cont}}}
\def\Paalpha{\mbox{Pa$\hspace{0.2ex}\alpha$}}
\def\Bralpha{\mbox{Br$\hspace{0.2ex}\alpha$}}

\def\NaD{\mbox{Na\,\specchar{i}\,D}}    
\def\NaDone{\mbox{Na\,\specchar{i}\,\,D$_1$}}
\def\NaDtwo{\mbox{Na\,\specchar{i}\,\,D$_2$}}
\def\NaID{\mbox{Na\,\specchar{i}\,\,D}}
\def\NaIDone{\mbox{Na\,\specchar{i}\,\,D$_1$}}
\def\NaIDtwo{\mbox{Na\,\specchar{i}\,\,D$_2$}}
\def\Done{\mbox{D$_1$}}
\def\Dtwo{\mbox{D$_2$}}

\def\Mgbone{\mbox{Mg\,\specchar{i}\,b$_1$}}
\def\Mgbtwo{\mbox{Mg\,\specchar{i}\,b$_2$}}
\def\Mgbthree{\mbox{Mg\,\specchar{i}\,b$_3$}}
\def\MgIb{\mbox{Mg\,\specchar{i}\,b}}
\def\MgIbone{\mbox{Mg\,\specchar{i}\,b$_1$}}
\def\MgIbtwo{\mbox{Mg\,\specchar{i}\,b$_2$}}
\def\MgIbthree{\mbox{Mg\,\specchar{i}\,b$_3$}}

\def\CaIIK{\mbox{Ca\,\specchar{ii}\,K}}       
\def\CaIIH{\mbox{Ca\,\specchar{ii}\,H}}
\def\CaIIHK{\mbox{Ca\,\specchar{ii}\,H\,\&\,K}}
\def\HK{\mbox{H\,\&\,K}}
\def\Kthree{\mbox{K$_3$}}      
\def\Hthree{\mbox{H$_3$}}
\def\Ktwo{\mbox{K$_2$}}
\def\Htwo{\mbox{H$_2$}}
\def\Kone{\mbox{K$_1$}}     
\def\Hone{\mbox{H$_1$}}     
\def\KtwoV{\mbox{K$_{2V}$}}
\def\KtwoR{\mbox{K$_{2R}$}}
\def\KoneV{\mbox{K$_{1V}$}}
\def\KoneR{\mbox{K$_{1R}$}}
\def\HtwoV{\mbox{H$_{2V}$}}
\def\HtwoR{\mbox{H$_{2R}$}}
\def\HoneV{\mbox{H$_{1V}$}}
\def\HoneR{\mbox{H$_{1R}$}}

\def\hk{\mbox{h\,\&\,k}}
\def\kthree{\mbox{k$_3$}}    
\def\hthree{\mbox{h$_3$}}
\def\ktwo{\mbox{k$_2$}}
\def\htwo{\mbox{h$_2$}}
\def\kone{\mbox{k$_1$}}     
\def\hone{\mbox{h$_1$}}     
\def\ktwoV{\mbox{k$_{2V}$}}
\def\ktwoR{\mbox{k$_{2R}$}}
\def\koneV{\mbox{k$_{1V}$}}
\def\koneR{\mbox{k$_{1R}$}}
\def\htwoV{\mbox{h$_{2V}$}}
\def\htwoR{\mbox{h$_{2R}$}}
\def\honeV{\mbox{h$_{1V}$}}
\def\honeR{\mbox{h$_{1R}$}}

\ifnum\preprintheader=1     
\makeatletter  
\def\@maketitle{\newpage
\markboth{}{}%
  {\mbox{} \vspace*{-8ex} \par 
   \em \footnotesize To appear in ``Magnetic Coupling between the Interior 
       and the Atmosphere of the Sun'', eds. S.~S.~Hasan and R.~J.~Rutten, 
       Astrophysics and Space Science Proceedings, Springer, 
       Heidelberg, 2009.} \vspace*{-5ex} \par
 \def\lastand{\ifnum\value{@inst}=2\relax
                 \unskip{} \andname\
              \else
                 \unskip \lastandname\
              \fi}%
 \def\and{\stepcounter{@auth}\relax
          \ifnum\value{@auth}=\value{@inst}%
             \lastand
          \else
             \unskip,
          \fi}%
  \raggedright
 {\Large \bfseries\boldmath
  \pretolerance=10000
  \let\\=\newline
  \raggedright
  \hyphenpenalty \@M
  \interlinepenalty \@M
  \if@numart
     \chap@hangfrom{}
  \else
     \chap@hangfrom{\thechapter\thechapterend\hskip\betweenumberspace}
  \fi
  \ignorespaces
  \@title \par}\vskip .8cm
\if!\@subtitle!\else {\large \bfseries\boldmath
  \vskip -.65cm
  \pretolerance=10000
  \@subtitle \par}\vskip .8cm\fi
 \setbox0=\vbox{\setcounter{@auth}{1}\def\and{\stepcounter{@auth}}%
 \def\thanks##1{}\@author}%
 \global\value{@inst}=\value{@auth}%
 \global\value{auco}=\value{@auth}%
 \setcounter{@auth}{1}%
{\lineskip .5em
\noindent\ignorespaces
\@author\vskip.35cm}
 {\small\institutename\par}
 \ifdim\pagetotal>157\p@
     \vskip 11\p@
 \else
     \@tempdima=168\p@\advance\@tempdima by-\pagetotal
     \vskip\@tempdima
 \fi
}
\makeatother     
\fi

\title*{The Evershed Flow and the Brightness of the Penumbra}

\author{L. R. Bellot Rubio}

\authorindex{Bellot Rubio, L. R.} 

\institute{Instituto de Astrof\'{\i}sica de Andaluc\'{\i}a, CSIC, Granada, Spain}

\maketitle

\setcounter{footnote}{0}  

\begin{abstract} 
The Evershed flow is a systematic motion of gas that occurs in the
penumbra of all sunspots. Discovered in 1909, it still lacks a
satisfactory explanation. We know that the flow is magnetized, often
supersonic, and that it shows conspicuous fine structure on spatial
scales of 0.2\arcsec\/--0.3\arcsec\/, but its origin remains
unclear. The hope is that a good observational understanding of the
relation between the flow and the penumbral magnetic field will help
us determine its nature. Here I review advances in the
characterization of the Evershed flow and sunspot magnetic fields from
high-resolution spectroscopic and spectropolarimetric
measurements. Using this information as input for 2D heat transfer
simulations, it has been demonstrated that hot Evershed upflows along
nearly horizontal field lines are capable of explaining one of the
most intriguing aspects of sunspots: the surplus brightness of the
penumbra relative to the umbra. They also explain the existence of
penumbral filaments with dark cores. These results support the idea
that the Evershed flow is largely responsible for the transport
of energy in the penumbra.
\end{abstract}

\section{Introduction}     

The Evershed effect was discovered a century ago as a Doppler 
shift of spectral lines in sunspots away from disk center
\citep{bellot-1909MNRAS..69..454E}. The shift is to the red 
in the limb-side penumbra and to the blue in the center-side
penumbra, and happens together with strong line asymmetries.
\citep[e.g.,][]{bellot-1980A&A....82..157S}. Already from
the very beginning, these signatures were explained in terms of a
nearly horizontal outflow of gas -- the Evershed flow. The motions
that give rise to the Evershed effect represent the most conspicuous
dynamical phenomenon observed in sunspots. Advances in instrumentation
have allowed us to characterize them with increasing degree of detail,
but their origin remains a fundamental problem in sunspot
physics. 

Another unsolved problem is the brightness of the penumbra, which
radiates about 75\% of the quiet Sun flux. This is significantly
larger than the 20\% emitted by the umbra. In recent years, 
considerable efforts have been made to understand the energy transport
in sunspots. \citet{bellot-1994A&A...290..295J} proposed convection by
interchange of magnetic flux tubes as the process that heats the
penumbra, and \citet{bellot-1998A&A...337..897S} developed the moving
tube model to simulate it. This mechanism was later found to be unable
to explain the brightness of the penumbra \citep{bellot-2003A&A...411..249S, 
bellot-2003A&A...411..257S}, but the moving tube model 
survived as a reasonably good, yet idealized, description 
of the penumbral magnetic field.

In the photosphere, the penumbra is a conglomerate of nearly 
horizontal field lines embedded in a stronger and more vertical field,
as first proposed by \citet{bellot-1993A&A...275..283S} in their uncombed
model. Strong gradients or discontinuities of the atmospheric
parameters occur along the line of sight due to the vertical
interlacing of different magnetic components. This characteristic
organization of the field explains many features of the penumbra,
including its filamentary appearance in continuum intensity and
polarized light (see \cite{bellot-2003ASPC..286..211S};
\cite{bellot-2007hsa..conf..271B}; \cite{bellot-2008arXiv0810.0080B}, 
\cite{bellot-2008arXiv0811.2747S}; and \cite{bellot-tritschler09} for reviews). 

The Evershed flow occurs along the more horizontal field lines and is
a global phenomenon, so it represents an excellent candidate to heat
the penumbra. Based on the radiative cooling times of hot plasma
embedded in a non-stratified atmosphere, \citet{bellot-2003A&A...411..257S}
concluded that the Evershed flow would be able to provide the required
amount of energy if the penumbral field lines return to the solar
surface soon after they emerge into the photosphere. At that time, the
existence of such field lines was unclear and many considered this
result as a serious difficulty for the otherwise successful uncombed
model.

The direct detection of small patches of magnetic fields diving back
below the solar surface everywhere in the mid and outer penumbra
\citep{bellot-2007PASJ...59S.593I, bellot-2008A&A...481L..21S} 
removes this problem. Also, the radiative cooling times in a
stratified atmosphere which is continuously heated by the Evershed
flow may be quite different from those estimated by Schlichenmaier et
al.\ (1999). With longer cooling times, a single flow channel would be
able to heat larger penumbral areas than previously thought. These two
considerations call for a re-examination of the Evershed flow as the
mechanism responsible for the brightness of the penumbra.

Here I summarize the properties of the Evershed flow as deduced from
high-resolution spectroscopic and spectropolarimetric measurements
(Section~\ref{bellot-properties}). Any serious attempt to model the
energy transport in the penumbra has to include this
information. \citet{bellot-2008A&A...488..749R} used it to solve the
heat transfer equation in a stratified atmosphere consisting of hot
Evershed flows along nearly horizontal magnetic flux tubes. The
calculations indicate that the tubes would be observed as bright
penumbral filaments with a central dark lane. The filaments are heated
by the flow over a length of about 3000~km; together with their
surroundings, they emit 50\% of the the quiet Sun intensity. These
results suggest that the surplus brightness of the penumbra is a
natural consequence of the Evershed flow
(Section~\ref{bellot-heattransfer}). Another mechanism proposed to
explain the brightness of the penumbra is overturning convection
(Section~\ref{bellot-overturning}). However, this idea will remain
speculative until measurements at 0.1\arcsec\ demonstrate the
existence of convective motions in penumbral filaments.

\section{The Evershed Flow at High Spatial Resolution} 
\label{bellot-properties}

Most of what we know about the Evershed flow has been learned from
high-resolution filtergrams and spectropolarimetric measurements at
1\arcsec\/--2\arcsec\ \citep[see][ and references
therein]{bellot-tritschler09}. Despite their moderate angular
resolution, the latter have resulted in a very detailed
characterization of the flow thanks to its relation with the more
inclined fields of the penumbra, which allows inversion techniques to
separate the different magnetic atmospheres that coexist in the pixel.

With the advent of instruments capable of reaching 0.3\arcsec, most
notably the Universal Birefringent Filter at the Dunn Solar Telescope,
the TRIPPEL spectrograph at the Swedish Solar Telescope, and the
spectropolarimeter of the Solar Optical Telescope aboard Hinode, the
results derived from Stokes inversions have been confirmed and often
extended in a more direct way. As a consequence we now have a rather
good understanding of the Evershed flow. In the next paragraphs I 
will discuss the properties of the flow at high spatial resolution 
and its relation with the magnetic field of the penumbra based on these
observations.

\subsubsection{The Evershed Flow Occurs in the Dark Cores of Penumbral 
Filaments}

Bright filaments in the inner penumbra consist of a central dark core
surrounded by two lateral brightenings
\citep{bellot-2002Natur.420..151S, bellot-2004A&A...424.1049S}. The
distance between the lateral brightenings is typically 100 km, but in
some cases it may be as large as 300~km, making these structures easy
targets for 50 cm telescopes.

\citet{bellot-2005A&A...443L...7B} performed spectroscopy of dark-cored
penumbral filaments with the Swedish Solar Telescope. They used the
TRIPPEL spectrograph to observe the \FeI~557.6~nm and 709.0~nm
lines at a resolution of 0.2\arcsec. A bisector analysis of the data
revealed that the Evershed flow is stronger in the dark cores,
although the lateral brightenings also show non-negligible
velocities. The velocities were found to increase with depth in the
photosphere. The observed variation of the Doppler shift with position
in the penumbra and heliocentric angle indicated that the Evershed
flow is directed {\em upward\/} in the inner penumbra. Also, the Zeeman
splitting of the \FeII~614.9~nm line was consistent with
slightly weaker fields in the dark cores as compared with the lateral
brightenings or the surrounding medium.

The observation that the Evershed flow occurs preferentially in the
dark cores of penumbral filaments has been confirmed by
\citet{bellot-2005A&A...436.1087L, bellot-2007A&A...464..763L}, 
\citet{bellot-2006ApJ...646..593R}, and \citet{bellot-2008ApJ...672..684R} 
using high-resolution filtergrams. The \FeI\ 630.25~nm magnetograms of
\citet{bellot-2005A&A...436.1087L} and
\citet{bellot-2008A&A...489..429V} show the dark cores with weaker 
signals than the bright edges, suggesting more inclined fields. The
field could also be weaker, but no definite conclusion can be made
because of possible saturation effects (in the strong field regime,
the amplitude of the circular polarization signal does not increase
with the field strength). Interestingly, these measurements do not
show negative polarities across the filaments, that is, the field in
the dark cores and the lateral brightenings appears to have the same
orientation (but see the remarks by \cite{bellot-sanchezalmeida09}).  A
similar conclusion has been reached by
\citet{bellot-2007ApJ...668L..91B} from an analysis of Hinode
spectropolarimetric measurements: the inclination of the magnetic
field is slightly larger in the dark cores, but no change of polarity
occurs with respect to the surroundings (at least at a resolution of
0.3\arcsec\/).

An intriguing feature revealed by high-resolution Dopplergrams 
(Hirzber\-ger \& Kneer 2001; \cite{bellot-2002A&A...389.1020R, 
bellot-2006A&A...453.1117B}) and spectropolarimetric measurements 
(\cite{bellot-2004A&A...427..319B, bellot-2008A&A...481L...9I}) is that 
there is no place in the penumbra where the velocities drop to zero, 
except where they should vanish because of projection effects. For example, 
in the inner penumbra the stronger flows occur in the bright filaments, but
smaller velocities are also seen in between them. The origin of these
motions remains unexplained.

\subsubsection{The Evershed Flow Returns to the Solar Surface in the Middle 
Penumbra and Beyond} 

The large-scale geometry of the flow can be inferred from the
azimuthal variation of the line-of-sight (LOS) velocity as a function
of radial distance. The results of this analysis indicate that, on
average, the flow is slightly inclined upward in the inner penumbra,
then becomes horizontal at about 0.8 penumbral radii, and finally
returns to the solar surface in the middle and outer penumbra with
inclinations of 10$^{\circ}$--30$^{\circ}$ to the horizontal
(\cite{bellot-2000A&A...358.1122S, bellot-2003A&A...403L..47B,
bellot-2004A&A...415..717T, bellot-2006A&A...453.1117B}, and
references therein). This must be understood as an average behavior;
of course, upflows also exist in the outer penumbra, but the downflows
dominate the azimuthal average.

\begin{figure}[t]
\begin{center}
\resizebox{!}{.49\hsize}{\includegraphics[bb=119 400 745 1023]{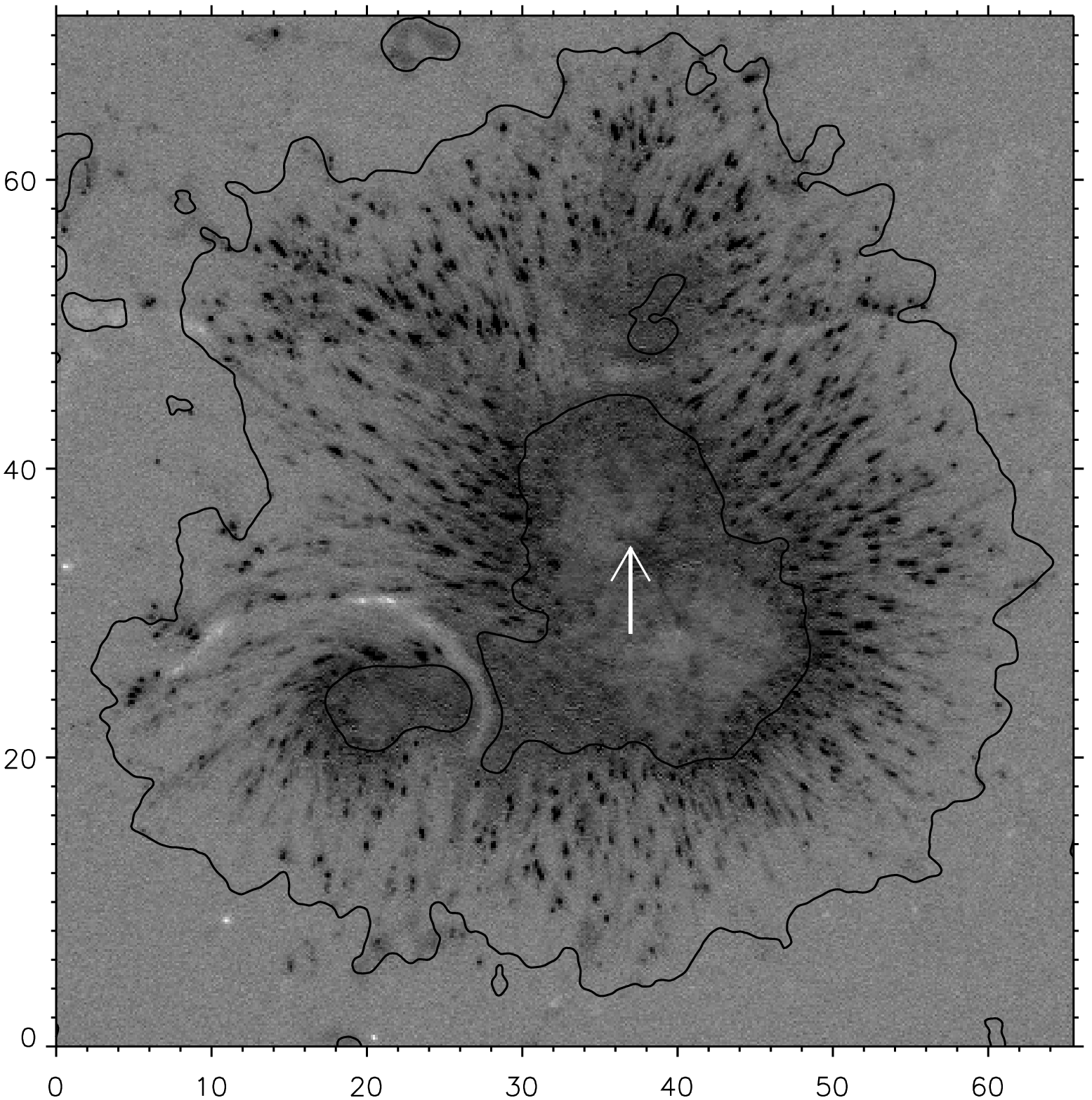} }
\resizebox{!}{.49\hsize}{\includegraphics[bb=119 400 745 1023]{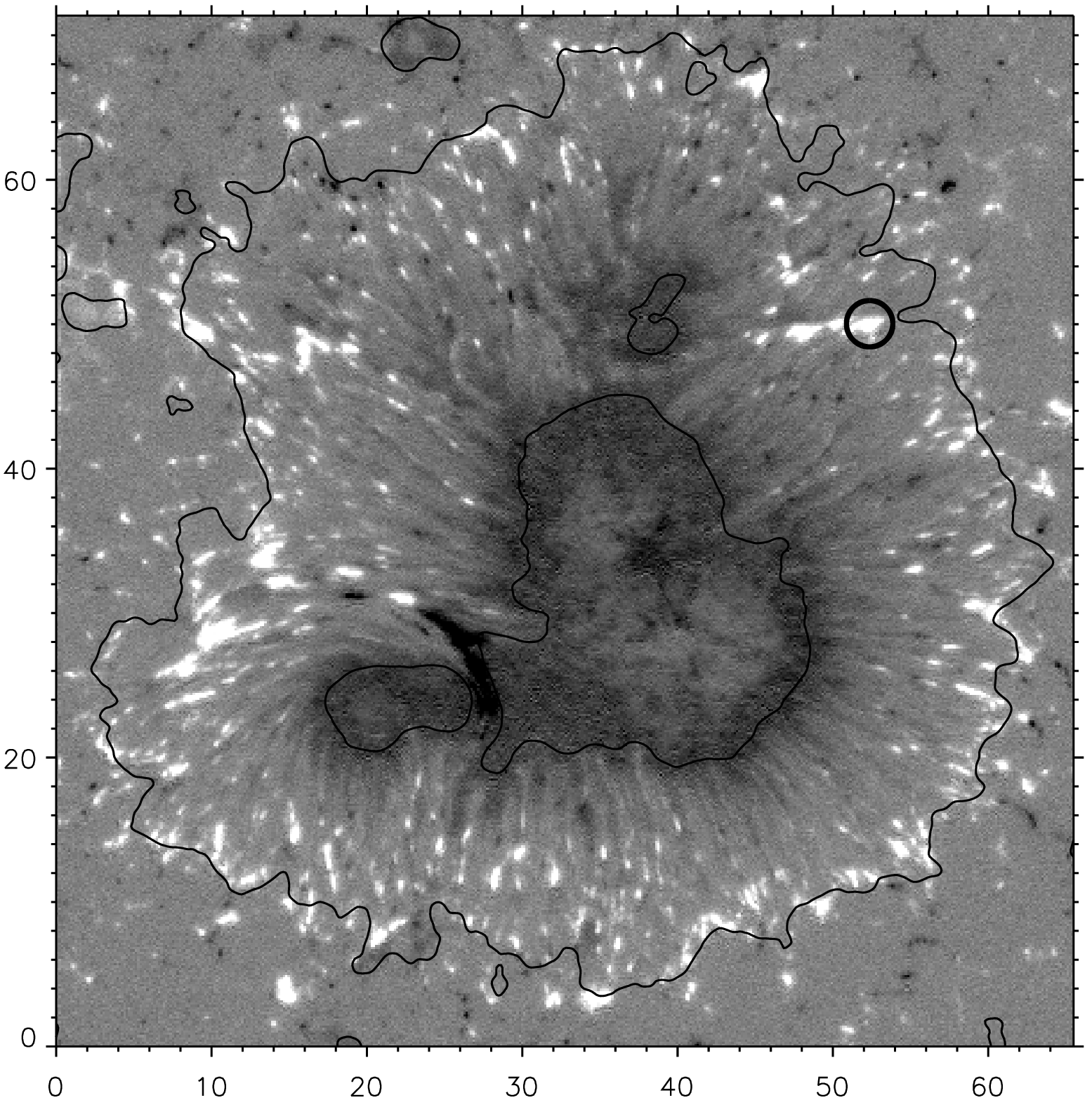} }
\end{center}
\caption[]{Stokes $V$ maps of AR 10973 at $-27.7$~pm ({\em left\/}) 
and $+27.7$~pm ({\em right\/}) from the center of the \FeI~630.25~nm
line. The data were taken by the Hinode spectropolarimeter on 1 May
2007, with the spot at a heliocentric angle of 6$^\circ$.  {\em Black\/}
represents the polarity of the spot in the two maps. The {\em circle\/}
overplotted in the {\em right\/} panel marks the downflowing patch considered
in Fig.~\ref{bellot-fig2}. Distances are given in arcsec, and the
{\em arrow\/} points to disk center.}
\label{bellot-fig1}
\end{figure}

On smaller scales, \citet{bellot-2006ApJ...646..593R} employed
\FeI~557.6~nm filtergrams taken at the Dunn Solar Telescope to
demonstrate that the Evershed flow emerges as a hot upflow in bright
penumbral grains and quickly becomes horizontal along individual
filaments. This result has been confirmed by
\citet{bellot-2007PASJ...59S.593I}, who had the ingenious idea of
using Stokes $V$ maps in the far wings of \FeI\ 630.25~nm to identify
strongly blueshifted or redshifted polarization
signals. Figure~\ref{bellot-fig1} shows two such magnetograms for AR
10973 at $-27.7$~pm and $+27.7$~pm from line center, as observed by
Hinode on 1 May 2007. The heliocentric angle of the spot
was~6$^\circ$. The signs have been reversed in the red-wing
magnetogram so that black indicates the same polarity in the two
panels. The structures visible in Fig.~\ref{bellot-fig1} represent
strongly Doppler-shifted polarization signals. However, the strength
of the signal does not inform about the velocity itself, as it also
depends on other parameters such as the field inclination or the gas
temperature.

Figure~\ref{bellot-fig1} shows with unprecedented clarity the sources
and sinks of the Evershed flow (for additional information, see the
paper by \cite{bellot-ichimoto09}). The blue wing magnetogram is
dominated by small patches of strong upflows located preferentially
(but not exclusively) in the inner penumbra. The red wing magnetogram
shows similar patches of downflows scattered all over the middle and
outer penumbra. Strong downflows occur in the inner penumbra too, but
they are less frequent. \citet{bellot-1997Natur.389...47W} were the
first to detect downflows in the penumbra, near the sunspot edge. The
polarity of the magnetic signals in the downflowing patches is
opposite to that of the spot. Thus, not only the flow, but also the
magnetic field, returns to the solar surface.  Observed at lower
spatial resolution, this pattern of upflows/downflows would result in
the average radial behavior mentioned above.

The flow field revealed by the Hinode magnetograms is highly
organized. Presumably each upflow in the inner penumbra connects with
a downflow in the mid or outer penumbra, although this needs to be
verified by studying the evolution of the flow. The patches are larger
than the angular resolution of the Hinode spectropolarimeter
(0.3\arcsec), suggesting that they are resolved structures. Moreover,
the downflows occur intermittently, not in every pixel as indicated by
the inversions of \citet{bellot-2005ApJ...622.1292S}. This is
consistent with the idea that the Evershed flow is confined to
discrete channels oriented radially; the far-wing magnetograms only
show the relatively vertical inner and outer footpoints of these
structures. The channels seem to be shorter than the width of the
penumbra, although it is likely that their lengths vary as they evolve
with time.

\subsubsection{The Evershed Flow is Often Supersonic}

The Evershed flow attains the largest velocities in the outer
penumbra. Spectropolarimetry at 1\arcsec\/ has shown that the
azimuthally averaged flow is nearly supersonic from the middle
penumbra outward \citep{bellot-2004A&A...427..319B}. There have been reports
of strongly Doppler-shifted line satellites in the penumbra
\citep{bellot-1995A&A...298L..17W}, and numerical models of the Evershed flow
also predict supersonic velocities \citep{bellot-1997Natur.390..485M,
bellot-1998A&A...337..897S}. However, these velocities have not been detected
directly until the advent of Hinode.

\begin{figure}[t]
\begin{center}
\resizebox{!}{.34\hsize}{\includegraphics[bb=54 380 564 633]{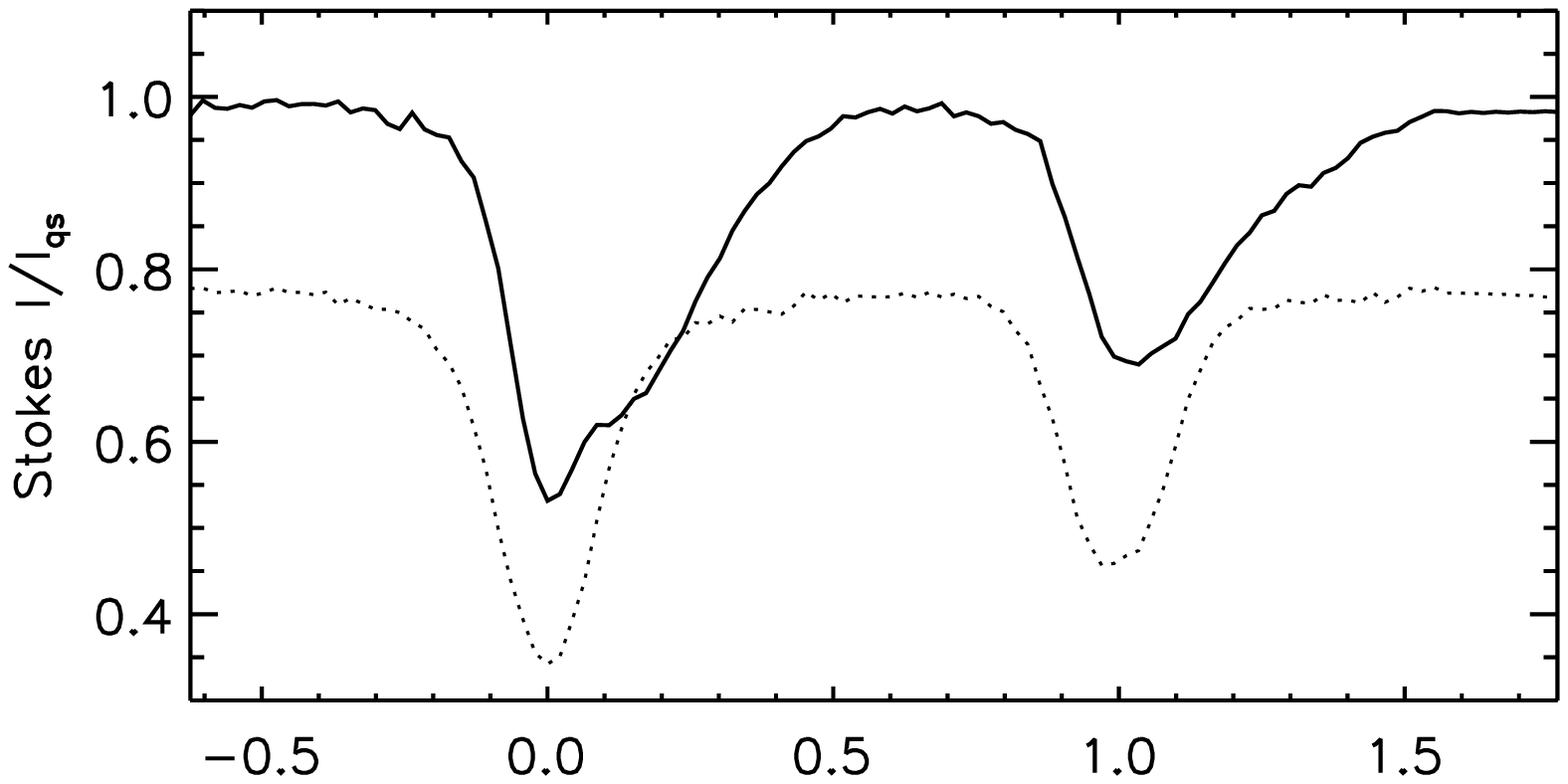} }
\resizebox{!}{.34\hsize}{\includegraphics[bb=54 380 564 633]{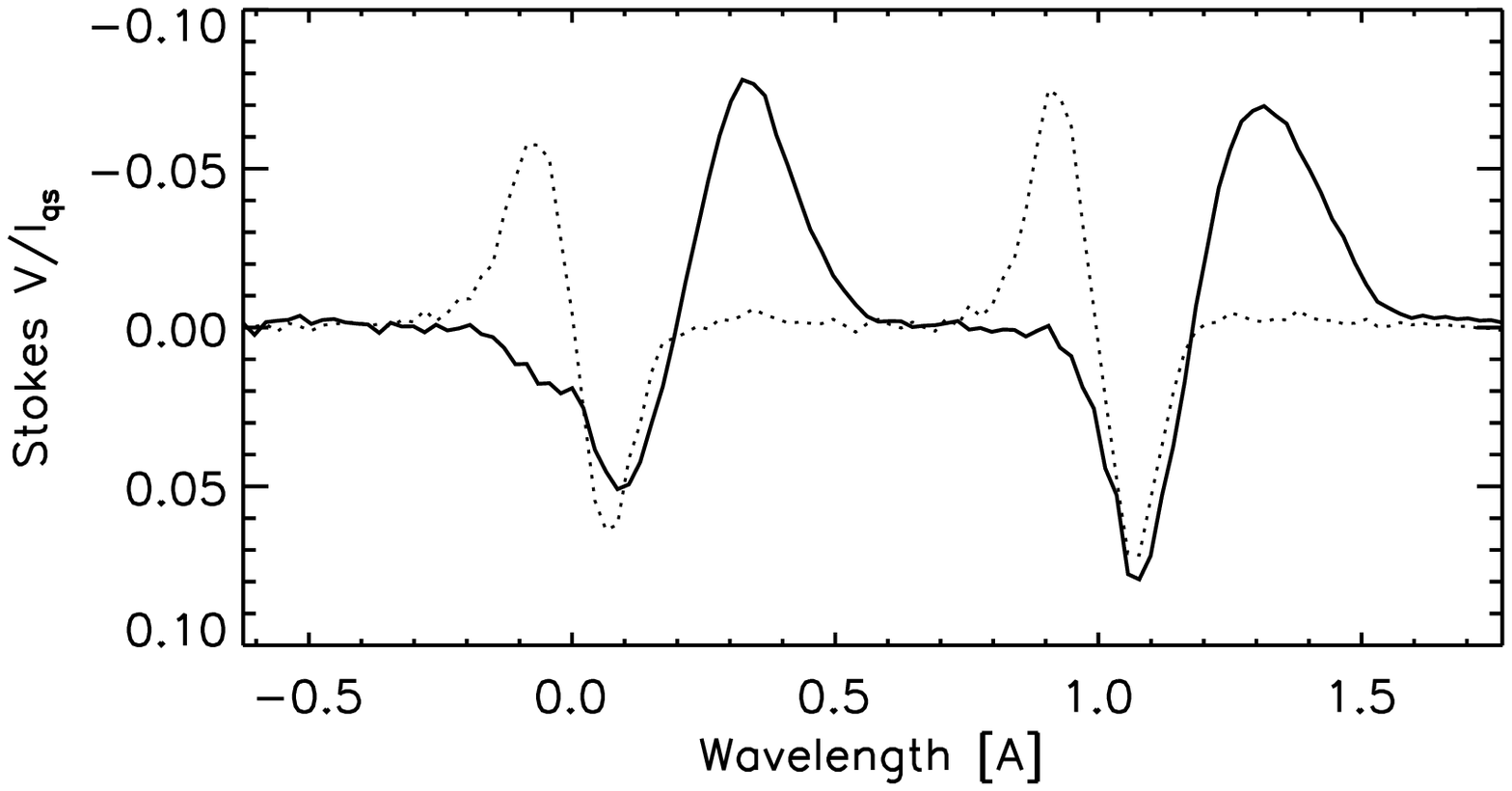} }
\end{center}
\caption[]{Stokes $I$ and $V$ profiles observed in one of the 
downflowing patches of Fig.~\ref{bellot-fig1} and a nearby pixel at
the same radial distance ({\em solid\/} and {\em dotted\/} lines,
respectively). The exact position of the profiles is indicated in the
{\em right\/} panel of Fig.~\ref{bellot-fig1} with a {\em
circle\/}. The {\em dotted\/} lines represent typical signals in the outer
penumbra, with the polarity of the spot and very small flows (the
observations were taken close to disk center). The zero of the
wavelength scale corresponds to the line-core position of the average
quiet Sun intensity profile, computed from pixels with Stokes $V$
amplitudes smaller than 0.5\% of the continuum intensity.}
\label{bellot-fig2}
\end{figure}

Figure~\ref{bellot-fig2} displays the Stokes spectra observed in one
of the downflowing patches of Fig.~\ref{bellot-fig1} and a nearby
pixel used as a reference (solid and dotted lines, respectively). The
spectra represented by the solid lines are remarkable in a number of
ways. To start with, the intensity profiles exhibit a very bright
continuum and strongly tilted red wings. The tilt is maximum near the
continuum, indicating that the flow velocity increases with depth in
the atmosphere. The Stokes $V$ profiles are normal and have two
lobes. However, they are redshifted as a whole by 19.5~pm or
$\sim$9~km~s$^{-1}$. This LOS velocity is smaller than the true
velocity because of projection effects, but it already exceeds the
sound speed. The clear signatures of supersonic velocities in the
spectra recorded by Hinode are a consequence of the high spatial
resolution, which makes it possible to separate the flow channels from
their surroundings.

\begin{figure}[t]
\begin{center}
\resizebox{!}{.73\hsize}{\includegraphics[bb=60 56 409 400]{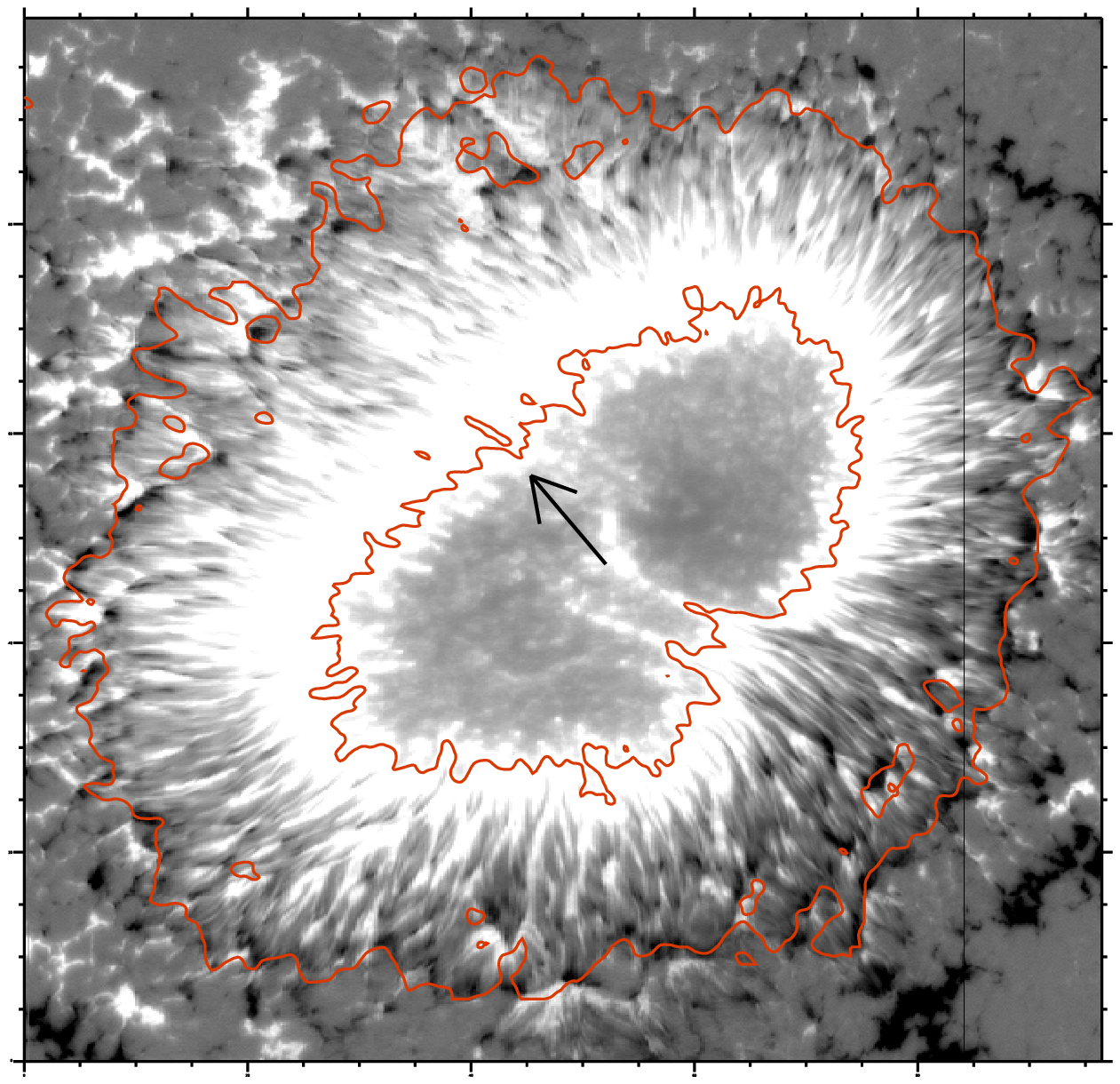} }
\resizebox{!}{.73\hsize}{\includegraphics[bb=55 56 165 400]{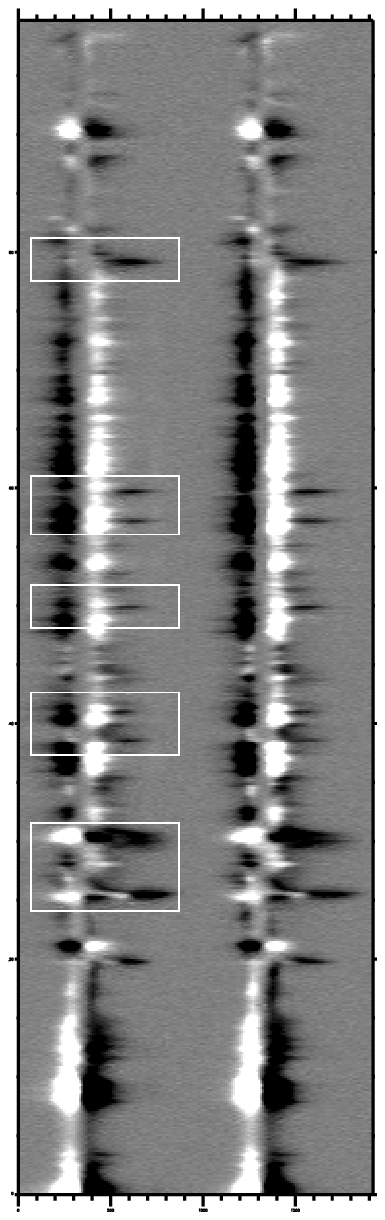} }
\end{center}
\caption[]{{\em Left:\/} AR 10923 as observed with 
the Hinode spectropolarimeter on 14 November 2006 between 16:30 and
17:16 UT. The map shows the integral of the Stokes $V$ profile of
\FeI\ 630.25~nm over the red lobe. {\em Black\/} and {\em white} 
represent opposite magnetic polarities. The {\em arrow} indicates the
direction to disk center. The heliocentric angle of the spot was
$8^\circ$. {\em Right:\/} Circular polarization profiles of \FeI\
630.15 and 630.25~nm along the slit indicated in the {\em left}
panel. The {\em rectangles} show examples of pixels with opposite
polarities and very strong Doppler shifts, both at the outer penumbral
boundary and well within the penumbra.}
\label{bellot-fig3}
\end{figure}

In general, the downflowing patches observed in the middle and outer
penumbra exhibit anomalous Stokes $V$ profiles with three lobes
(Fig.~\ref{bellot-fig3}). They can be interpreted as the superposition
of two regular signals. One of them has the polarity of the spot and
is relatively unshifted, while the other has opposite polarity and
shows a very strong redshift. The two signals do not necessarily come
from different atmospheres in the pixel; at the resolution of Hinode,
it is more likely that they are produced by two magnetic components
stacked along the line of sight. We know that this configuration must
exist in the penumbra because the observed Stokes $V$ profiles have
nonzero area asymmetries. The inversions of
\citet{bellot-2007PASJ...59S.601J} also demonstrate that models with
strong gradients in the vertical direction are able to reproduce the
complex profiles measured by Hinode without any horizontal 
interlacing of different magnetic components.

\subsubsection{The Evershed Flow is Associated with the Weaker and
More Inclined Fields of the Penumbra} 

A well-established observational result is that the Evershed flow
happens along the more horizontal fields of the penumbra (see
\cite{bellot-tritschler09} for a review). Indeed, it has been demonstrated
that the azimuthally averaged flow is parallel to the magnetic field
vector all the way from the inner to the outer penumbra
\citep{bellot-2003A&A...403L..47B}. Near the umbra the inclined fields 
reside in the bright filaments \citep[e.g.,][]{bellot-2007PASJ...59S.601J},
but at larger radial distances they tend to occur in dark
structures.  It is also accepted that the flow is associated with the
weaker fields of the penumbra, except perhaps in the outer penumbra
where the strength of the ambient field decreases rapidly
\citep{bellot-2004A&A...427..319B, bellot-2005A&A...436..333B, 
bellot-2006A&A...450..383B, bellot-2007ApJ...671L..85T, 
bellot-2008A&A...480..825B}.

For the most part, these results have been obtained from the inversion
of spectropolarimetric measurements that do not spatially resolve the
different components of the penumbra. \citet{bellot-1993ApJ...403..780T},
\citet{bellot-1995A&A...298..260R}, \citet{bellot-2005A&A...436.1087L}, and 
others investigated the relation between the flow and the magnetic 
field using high-resolution magnetograms where penumbral filaments 
are clearly distinguished, but their results cannot be considered
complete because of the lack of linear polarization measurements.

\begin{figure}[t]
\begin{center}
\resizebox{!}{.749\hsize}{\includegraphics{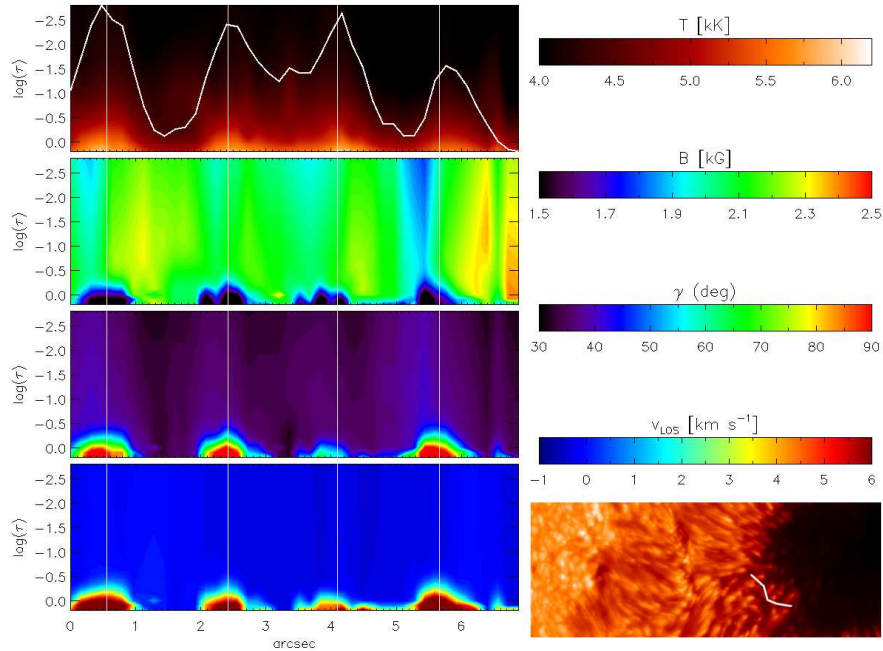} }
\end{center}
\caption[]{Vertical stratification of atmospheric parameters along
a cut in the inner penumbra of AR 10973 as observed by Hinode on 10
May 2006, at $\theta = 46^\circ$. The cut crosses four bright
filaments of the limb-side penumbra ({\em lower right corner\/} of the
figure). The {\em left} panels show, from {\em top\/} to {\em
bottom\/}, temperature, field strength, field inclination (in the LOS
reference frame), and LOS velocity as a function of optical depth,
with {\em color bars\/} to the {\em right}. The curve overplotted in
the temperature panel represents the continuum intensity along the
cut. See \citet{bellot-2007PASJ...59S.601J} for details. }
\label{bellot-fig4}
\end{figure}

Hinode has allowed us to study the magnetic and dynamic configuration
of individual filaments with full Stokes polarimetry and complex
atmospheric models. \citet{bellot-2007PASJ...59S.601J}, for instance,
applied an uncombed model consisting of a single magnetic atmosphere
with a flux tube occupying part of the line-forming region. Their
inversions provide the variation of the physical parameters with
height. Figure~\ref{bellot-fig4} shows a vertical cut crossing the
inner penumbra of AR 10923 on 10 November 2006. The cut samples four
bright penumbral filaments. As can be seen, the Evershed flow occurs
deep in the photosphere at the position of the filaments. It is
confined to narrow channels having weaker and more inclined
fields than the ambient medium. This is the first time that the magnetic
and thermodynamic structure of the flow channels is visualized
simultaneously in the vertical and horizontal directions with high
angular resolution. The results are less clear in the middle and outer
penumbra where it becomes difficult to trace individual filaments, but
the flow still seems to be associated with the more inclined fields
\citep{bellot-2008A&A...481L..17J}. Similar results have been derived
by \citet{bellot-2008A&A...481L..13B} from inversions of Hinode
measurements, confirming the overall picture of flow channels embedded
in a stronger and more vertical field. As expected, the ambient field
wraps around the flow channels, at least in the visible layers of the
photosphere
\citep{bellot-2008A&A...481L..13B}.

\subsubsection{The Evershed Flow is Magnetized}

It should be clear by now that the Evershed flow is magnetized, but
there are other arguments supporting this conclusion. One of them is
the net circular polarization (NCP) observed in the penumbra
\citep[e.g., ][]{bellot-1992ApJ...398..359S, bellot-2000A&A...361..734M,
bellot-2001ApJ...547.1130W, bellot-2002A&A...393..305M, 
bellot-2002A&A...381..668S, bellot-2007ApJ...671L..85T,
bellot-2008A&A...481L...9I}. Nonzero NCPs can be produced by flows of
field-free plasma, as happens, for instance, in the canopy of network
and facular magnetic elements. However, this mechanism creates
relatively small NCPs; the large values found in sunspots require the
flow to occur in a magnetized atmosphere.

Another argument is the existence of strong Doppler shifts in the
polarization profiles emerging from the penumbra. The shifts are
induced by the Evershed flow, which has to be magnetized to be able to
displace the Stokes signals. Figure~\ref{bellot-fig2} shows an extreme
example: it is just impossible to explain this kind of profiles with
field-free flows, because a magnetic field is required to generate the
observed polarization signals through the Zeeman effect. The
multi-lobed Stokes $V$ profiles occurring all over the penumbra (and
not only near the neutral line, as first pointed out by
\cite{bellot-2002NCimC..25..543B}) provide more examples of Doppler
shifts induced by flows of magnetized plasma. We have mentioned already 
that they can be interpreted as the superposition of two regular Stokes $V$ 
profiles, one of which is displaced in wavelength with respect to the 
other. Only flows occurring in a magnetized environment can shift 
the polarization signals to the extent required to generate 
Stokes $V$ spectra with three or more lobes (see, e.g., Fig.~1 
of Bellot Rubio 2006).

\citet{bellot-2008ApJ...687..668B} inverted Hinode measurements in terms 
of model atmospheres featuring vertical gradients of the
parameters. Their results show that the field strength of structures
associated with strong Evershed flows initially decreases with depth
in high photospheric layers, but then increases as the continuum
forming layers are approached. This behavior is incompatible with the
presence of field-free plasma near or just below $\tau =1$.

\subsubsection{The Evershed Flow Continues Beyond the Sunspot 
Border as MMFs} 

\citet{bellot-2006ApJ...649L..41C} have shown that some moving
magnetic features (MMFs) in the sunspot moat are generated by Evershed
clouds, that is, patches of enhanced Evershed flows that move radially
outward in the penumbra. Once in the moat, the Evershed clouds exhibit
large velocities for some time. These findings support the idea that
MMFs come from inside the spot \citep{bellot-2005ApJ...632.1176S,
bellot-2006SoPh..237..297R, bellot-2007ApJ...659..812K} and suggest
that the Evershed flow is not caused by small-scale convection in a
strong magnetic field, as it is also observed in the relatively
field-free environment of the moat.

\section{Heating of the Penumbra by the Evershed Flow}
\label{bellot-heattransfer}

\begin{figure}
\begin{center}
\resizebox{!}{.26\hsize}{\includegraphics[bb=13 10 340 147]{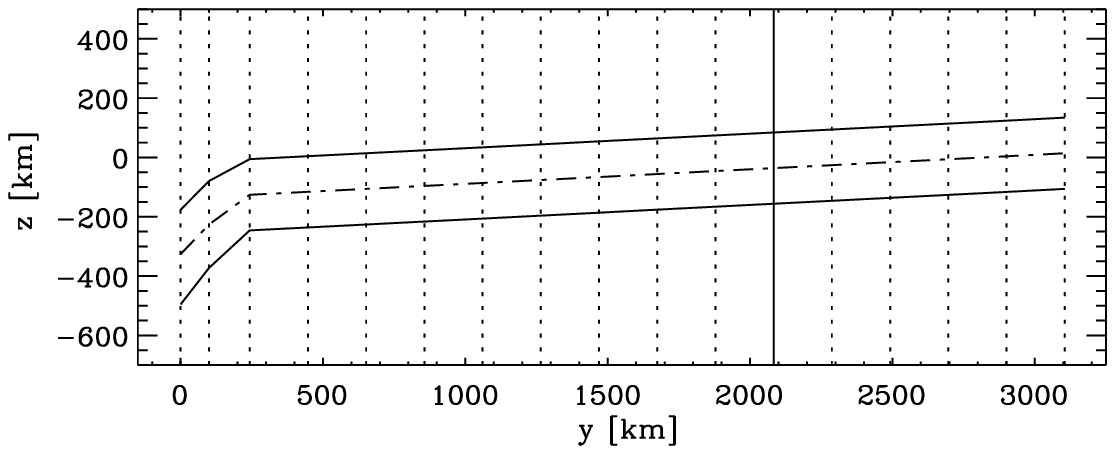}} 
\resizebox{!}{.27\hsize}{\includegraphics[bb= 15 9 263 197]{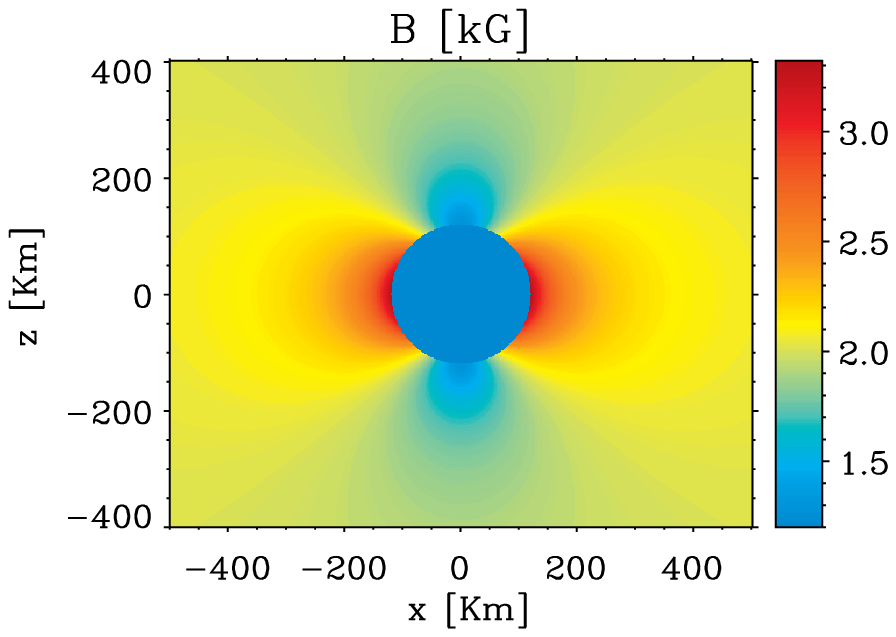}}
\end{center}
\caption[]{Magnetic flux tube embedded in a background (umbral-like) 
atmosphere. The tube has a diameter of 240~km. This configuration is
intended to represent bright filaments near the umbra/penumbra
boundary. {\em Left:\/} Vertical cut in the radial ($y$)
direction. The umbra is to the {\em left\/} and the outer penumbra to
the {\em right}. The vertical axis indicates height in the
photosphere. $z=0$~km corresponds to $\tau=1$. {\em Right:\/}
Cross-section of the tube. Displayed in color is the distribution of
the magnetic field strength. The tube has a weaker field than the
background atmosphere. However, the ambient field is potential and
therefore it shows strong spatial variations near the tube.  }
\label{bellot-fig5}
\end{figure}

In a recent paper, \citet{bellot-2008A&A...488..749R} investigated the
ability of the Evershed flow to explain the brightness of the
penumbra. To that end they solved the 2D heat transfer equation in a
simple model of a flow channel embedded in a stratified atmosphere
(Fig.~\ref{bellot-fig5}). Consistent with the results described above,
the simulated flow is radial, lies deep in the photosphere, and has
weaker and more inclined fields than the ambient medium. The flow
speed is 7~km~s$^{-1}$ at $y=0$~km, but it becomes supersonic farther
out.

\begin{figure}
\begin{center}
\resizebox{!}{.52\hsize}{\includegraphics[bb= -3 435 383 670]{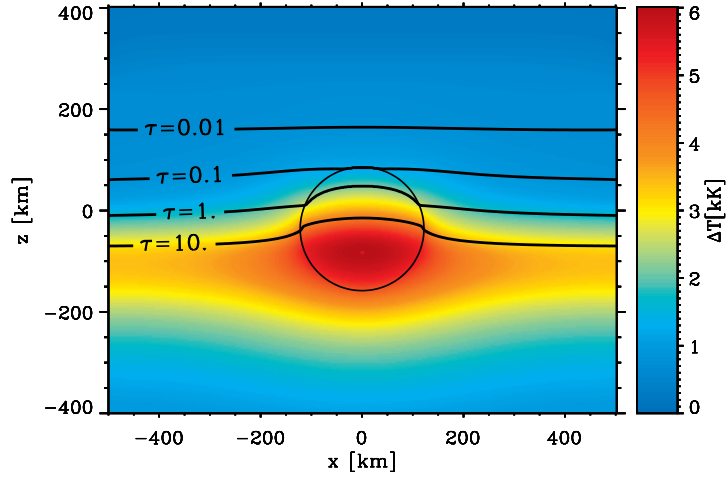}} 
\end{center}
\caption[]{Temperature perturbations induced by the Evershed flow in the 
$xz$-plane at $y=2083$~km (see Fig.~\ref{bellot-fig5}). The temperature
distribution without the flow is that of the cool umbral model of 
\citet{bellot-1994A&A...291..622C}. The {\em circle\/} 
represents the flow channel. The {\em solid lines\/} indicate constant
Rosseland optical depths. }
\label{bellot-fig6}
\end{figure}

The Evershed flow brings hot plasma from subsurface layers into the
photosphere, leading to temperature enhancements inside and outside
the flow channel. To determine the equilibrium temperature distribution,
the stationary heat transfer equation was solved numerically in 17
vertical planes spanning more than 3000~km in the radial direction. 
The temperatures obtained in this way ensure an exact balance between 
radiative losses and the energy supplied by the Evershed flow. As can 
be seen in Fig.~\ref{bellot-fig6}, the flow induces temperature 
enhancements of up to 6000~K. In the deepest photospheric layers 
accessible to the observations, around $\tau =1$, the temperature 
of the plasma is increased by nearly 3000~K.

\begin{figure}
\begin{center}
\resizebox{!}{.23\hsize}{\includegraphics[bb=  -10 0 273 77]{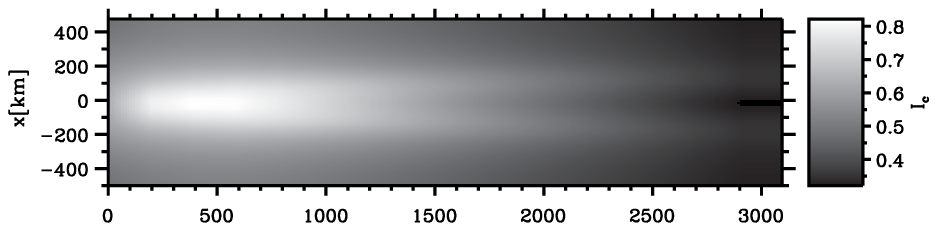}} 
\resizebox{!}{.23\hsize}{\includegraphics[bb=  -10 0 273 77]{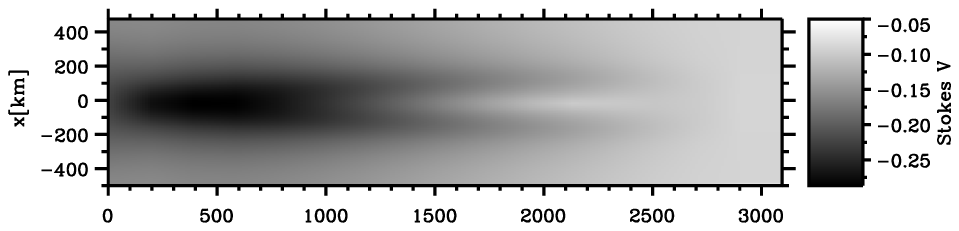}} 
\resizebox{!}{.23\hsize}{\includegraphics[bb=  -10 0 273 77]{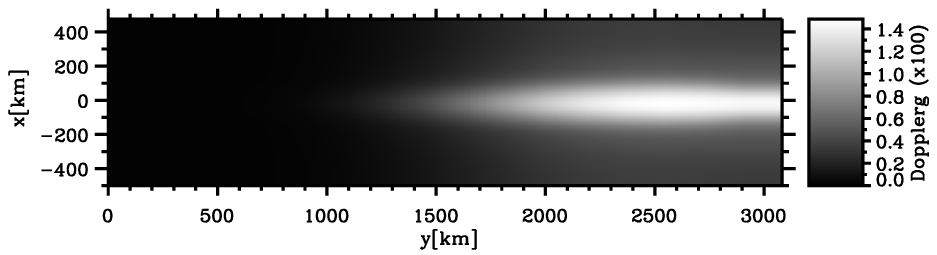}}
\end{center}
\caption[]{Magnetic flux tube of Fig.~\ref{bellot-fig5} as 
observed through a 1-m telescope in the \FeI~630.25~nm line. {\em
Top\/}: Continuum intensity at 630~nm. {\em Middle:\/} Stokes $V$ map
at $+10$~pm from line center. {\em Bottom:\/} Dopplergram calculated
using Stokes $I$ at $\pm 15$~pm from line center.}
\label{bellot-fig7}
\end{figure}

The physical parameters obtained from the calculations were used to
synthesize the Stokes profiles of the \FeI~630.25~nm line emerging
from the model. The results are presented in
Fig.~\ref{bellot-fig7}. The three panels display a continuum image, a
longitudinal magnetogram, and a Dopplergram. In continuum intensity,
the flow channel would be observed as a filament with a bright head
and a dark core. The lateral brightenings are separated by about
250~km. This filament is very similar to those observed in the inner
penumbra and has a length of 3000~km. The most important result,
however, is that the Evershed flow heats the surroundings very
efficiently: the average brightness in the box is approximately 
50\% of the quiet Sun value. Without the flow, it would be only 6\%,
corresponding to a very cool umbra. The average brightness is still a
bit low, but it should be possible to increase it with small changes
in the flow velocity and other simulation parameters.

The magnetogram of Fig.~\ref{bellot-fig7} shows weaker circular
polarization signal in the dark core as compared with the lateral
brightenings. This is exactly what is observed
(\cite{bellot-2005A&A...436.1087L, bellot-2007A&A...464..763L,
bellot-2007ApJ...668L..91B}; van Noort \& Rouppe van der Voort 
2008). Moreover, the
Dopplergram indicates that the dark core harbors the largest
velocities, in agreement with the findings of
\citet{bellot-2005A&A...443L...7B},
\citet{bellot-2005A&A...436.1087L}, and
\citet{bellot-2008ApJ...672..684R}.

A flow channel is capable of heating the atmosphere over a distance of
3000~km (Fig.~\ref{bellot-fig7}). Therefore, it does not need to
return to the solar surface within the short lengths estimated by
\citet{bellot-2003A&A...411..257S}. The fundamental difference here 
is a much longer cooling time due to the fact that, in the stationary 
phase, the background atmosphere has been heated by the flow and is 
already hot. Thus, the temperature gradient between the flow channel and the
surroundings is smaller than that considered by Schlichen\-maier et 
al.\ (1999). This leads to longer cooling times 
and longer filaments.

According to the calculations of \citet{bellot-2008A&A...488..749R},
bright penumbral filaments are the manifestation of deep-lying, radial
Evershed flows along nearly horizontal field lines. The dark cores are
produced by the weaker fields associated with the flows, which
increase the opacity and move the $\tau=1$ surface to higher layers
where the temperature is lower. The lateral brightenings represent the
walls of the flow channel and their immediate surroundings. Thus, the
dark core and the bright edges trace the same physical entity. This
explains why they follow parallel trajectories and move with the same
speed. The nonzero velocities observed in the bright edges can also
be understood in a natural way: they are produced by the Evershed
flow, but appear diluted because of the contribution of the ambient
medium, which is essentially at rest. The penumbral filaments are
deep-lying structures. The fact that the dark cores show up clearly 
in intensity measurements taken at the center of Zeeman-sensitive lines
\citep[e.g.,][]{bellot-2008ApJ...689L..69S} does not imply that they 
are formed high in the atmosphere. They may appear dark in line-core
images simply because they have weaker field strengths (which decreases
the separation between the $\sigma$-components of the Zeeman pattern)
and larger field inclinations (which increases the amplitude of the
central $\pi$-component). The two effect work in the same direction,
reducing the line-core intensity. The weaker and more inclined fields
can be located deep in the photosphere and still be able to modify the
line-core intensity due to the large width of the contribution
functions.

\section{Convection in Penumbral Filaments}
\label{bellot-overturning}

\citet{bellot-2006A&A...447..343S} and \citet{bellot-2006A&A...460..605S} 
proposed the idea of field-free gaps protruding into the sunspot
from below as an alternative mechanism to heat the penumbra. The gaps
would be located beneath the bright penumbral filaments and would
sustain overturning convection, carrying energy to the surface and
producing the Evershed flow (for a detailed explanation, see the
review by \cite{bellot-2008PhST..133a4015S}). Magnetoconvection
simulations of sunspots, while still in their infancy, appear to
support this picture \citep{bellot-2007ApJ...669.1390H}. The gappy
penumbral model is not free from problems, however, and radiative
transfer calculations are urgently needed to confront it with 
spectropolarimetric measurements \citep{bellot-2007hsa..conf..271B, 
bellot-2008arXiv0811.2747S, bellot-2008arXiv0810.0080B, bellot-thomas09}.

On the observational side, there have been some claims of the
detection of convective flows in the penumbra. Using Hinode filtergrams,
\citet{bellot-2007Sci...318.1597I} described a systematic twisting motion 
of brightness fluctuations in filaments located perpendicularly to the
line of symmetry. The motion always occurs from the limb toward
the observer, so the twist was interpreted as the manifestation of
overturning convection within the filaments. The Doppler measurements
of \citet{bellot-2008A&A...488L..17Z} and
\citet{bellot-2008ApJ...672..684R} indicate the existence of weak
upflows at the center of penumbral filaments and downflows on their
sides. As in normal granular convection, the upflows would turn into
downflows after releasing energy in the photosphere. This process of
heat transport in the vertical direction would explain the brightness
of the penumbra. The observed velocities are of the order of
1\,km~s$^{-1}$, well below those associated with the Evershed
flow. Interestingly, downflows at the periphery of penumbral filaments
are also present in the MHD simulations of
\citet{bellot-2009ApJ...691..640R}, albeit with larger velocities.

The question of whether overturning convection exists in the penumbra
is far from settled, however. The best spectroscopic measurements of
sunspots reach 0.2\arcsec, but they do not provide evidence for
overturning downflows in or around penumbral filaments, neither at
disk center nor at heliocentric angles of $\sim$20$^\circ$ (Fig.~2 of
\cite{bellot-2005A&A...443L...7B}). This is at odds with the
observations of \citet{bellot-2008A&A...488L..17Z} and
\citet{bellot-2008ApJ...672..684R}. To reach a definite conclusion we need
spectroscopic measurements at very high spatial resolution. With
0.1\arcsec\/ it should be possible to determine the flow field across
penumbral filaments, resolving internal fluctuations smaller than the
width of the filaments themselves. Hopefully, this kind of observations 
will be provided soon by instruments like IMaX aboard SUNRISE or 
CRISP at the Swedish Solar Telescope.

\section{Conclusions}
\label{bellot-conclusions}

The Evershed flow exhibits conspicuous fine structure at high angular
resolution. It occurs preferentially in the dark cores of penumbral
filaments, at least in the inner penumbra. The flow is magnetized and
often supersonic, as demonstrated by the observation of Stokes $V$
profiles shifted by up to 9~km~s$^{-1}$. At each radial distance, the
flow is associated with the more inclined fields of the penumbra; in
the inner penumbra this happens in the bright fi\-la\-ments, while in
the outer penumbra the dark filaments have the largest inclinations. 
The flow is also associated with weaker fields (except perhaps near 
the edge of the spot).

High-resolution magnetograms by Hinode show the sources and sinks of
the Evershed flow with unprecedented clarity, confirming earlier
results from Stokes inversions at lower resolution: on average, the
flow points upward in the inner penumbra, then becomes horizontal in
the middle penumbra, and finally dives down below the solar surface in
the outer penumbra. The Hinode observations reveal tiny patches of
upflows concentrated preferentially in the inner penumbra and patches
of downflows in the mid and outer penumbra; presumably they correspond
to the ends of individual flow channels.

Recent numerical calculations by \citet{bellot-2008A&A...488..749R}
have demonstrated that Evershed flows with these properties are
capable of heating the penumbra very efficiently, while reproducing
many other observational features such as the existence of dark-cored
penumbral filaments. This result strongly suggests that the radial
Evershed flow is indeed responsible for the brightness of the
penumbra.

At the same time, there have been observations of small-scale motions
in penumbral filaments that could reflect the existence of overturning
convection \citep{bellot-2007Sci...318.1597I, bellot-2008A&A...488L..17Z, 
bellot-2008ApJ...672..684R}. Convection is an essential ingredient of 
the field-free gap model proposed by \citet{bellot-2006A&A...447..343S} 
and seems to occur also in MHD simulations of sunspots 
\citep{bellot-2009ApJ...691..640R}. However, other spectroscopic 
observations at 0.2\arcsec\/ do not show clear evidence for downflows 
in filaments near the umbra/penumbra boundary 
\citep{bellot-2005A&A...443L...7B}. 

It is important to clarify whether or not convection exists in the
penumbra. To investigate this issue we need spectroscopic observations
at 0.1\arcsec\/. Narrow lanes of downflows should show up clearly in
those measurements. Only then will it be possible to assess the
contribution of overturning convection to the brightness of the
penumbra and compare it with that of the supersonic Evershed
flow. Ultimately, these efforts should reveal the primary mode of
energy transport in the penumbra. One possibility is that the two
mechanisms operate at the same time. In fact, the strong vertical
gradients of temperature observed within the flow channels
(Fig.~\ref{bellot-fig6}) may drive convective motions with upflows at
the center of the filaments and downflows along their sides. The gas
would not cross the field lines if the transverse component of the
field is similar to that displayed in Fig.~1 of
\citet{bellot-2007A&A...471..967B}. In any case, the velocities
associated with this process have to be small. The superposition of
the radial Evershed flow and overturning flows would result in two
helical, outward-twisted motions, one on each half of the
filament. The hot material would ascend at the center of the flow
channel while being displaced radially outward by the dominant
Evershed flow, and would descend along the filament edges after
releasing its energy in the photosphere. Even in that case, the
brightness of the penumbra would still be due to hot upflows channeled
by nearly horizontal field lines.

\begin{acknowledgement}
I thank Basilio Ruiz Cobo, Rolf Schlichenmaier, and Juan Manuel
Borrero for clarifying discussions on the role of the Evershed flow in
the heating of the penumbra. Financial support by the Spanish MICINN
through projects ESP2006-13030-C06-02 and PCI2006-A7-0624, and by
Junta de Andaluc\'{\i}a through project P07-TEP-2687 is gratefully
acknowledged.
\end{acknowledgement}

\begin{small}






\end{small}

\end{document}